\documentclass[graybox]{svmult}

\usepackage{mathptmx}       
\usepackage{helvet}         
\usepackage{courier}        
\usepackage{type1cm}        
\usepackage{makeidx}         
\usepackage{graphicx}        
\usepackage{multicol}        
\usepackage[bottom]{footmisc}
\usepackage{amsmath,dsfont}
\usepackage{amssymb}
\usepackage{epsfig}
\usepackage{calc}
\usepackage{amsfonts}
\usepackage{graphicx}
\usepackage[ansinew]{inputenc}
\usepackage{multirow}
\usepackage{subcaption}
\usepackage{tensor}
\usepackage[dvipsnames]{xcolor}
\usepackage{cancel}

\makeindex             

\begin{document}

\title*{Exact black hole solutions in higher-order scalar-tensor theories}
\author{Eugeny Babichev, Christos Charmousis and Nicolas Lecoeur}
\institute{Eugeny Babichev, \email{eugeny.babichev@ijclab.in2p3.fr}, Christos Charmousis, \email{christos.charmousis@ijclab.in2p3.fr}
\and Nicolas Lecoeur  \email{nicolas.lecoeur@ijclab.in2p3.fr}\at Universit\'e Paris-Saclay, CNRS/IN2P3,
IJCLab, 91405
Orsay, France}
\maketitle

\abstract{In this chapter, we discuss explicit black hole solutions in higher-order scalar-tensor theories. After a brief recap of no-hair theorems, we start our discussion by so-called stealth solutions present in theories with parity and shift symmetry. Stealth solutions are such that their metric are Ricci flat General Relativity solutions, but they are accompanied by a non-trivial scalar field, in both spherically-symmetric and rotating cases. The stealth metrics then enable to construct an analytic stationary solution of scalar-tensor theory which is called disformed Kerr metric. This solution constitutes a measurable departure from the usual Kerr geometry of GR. We discuss within parity and shift symmetric theories several non-stealth solutions. We then consider scalar-tensor theories stemming from a Kaluza-Klein reduction of a higher-dimensional Lovelock theory. These theories encompass all Horndeski functionals and hence go beyond parity and shift symmetry. Reduction and singular limits allow one to obtain non-stealth black holes with differing interesting properties which are not Ricci flat metrics. We analyse the solutions obtained and classify them with respect to the geometry of the internal space according to their Kaluza-Klein origin. }

\section{Black holes and scalar-tensor theories}
\label{sec:1}

General Relativity (GR) is a highly successful theory of gravity as it has predicted multiple gravitational phenomena and successfully passed multiple observational tests at different energy and length scales~\cite{Will:2014kxa}. It is an effective field theory (EFT) depending on a unique curvature scale. The puzzle of dark energy and dark matter, occurring at classical scales (IR) as well as the expected breakdown of GR at ultraviolet scales (UV) question the interval of validity of GR on both ends, at the IR and UV scales. It is important therefore to seek theories which consist measurable departures from the realm of GR and which are again effective theories (in the absence of a quantum theory of gravity). Gravity wave (GW) detections have put a very strong constraint on the speed of gravitons~\cite{LIGOScientific:2016aoc}, whose applicability however, for cosmological dark energy and higher-order scalar-tensor theories, is questionable~\cite{deRham:2018red}. Future GW detectors such as LISA~\cite{LISA:2017pwj} or the Einstein telescope~\cite{Maggiore:2019uih} will give better insight on constraints involving dark energy, and the cosmological puzzles remain open and are even enhanced as precision tests of cosmology improve. In the domain of strong gravity, black holes stand out as an extraordinary laboratory of strong gravity, and numerous ongoing observations such as gravitational waves of compact binaries~\cite{LIGOScientific:2016aoc}, the very large telescope interferometer of GRAVITY~\cite{GRAVITY:2018ofz} or again the Event Horizon Telescope \cite{EventHorizonTelescope:2020qrl}, will help to measure the validity and possible observational limits of GR. 

Many directions can be considered for modifying gravity (see~\cite{Clifton:2011jh} for a review), but geometric modifications all amount to altering the number of degrees of freedom as compared to the two metric degrees of freedom in GR. This follows from Lovelock's theorem~\cite{Lovelock:1971yv} which states that any four-dimensional action depending on the metric only and yielding second-order conserved field equations is the Einstein-Hilbert action with cosmological constant. The consequence is twofold: first, the most simple modification is to add a single degree of freedom, represented by a scalar field, second, many other modified gravity theories can be cast into a scalar-tensor form (see for instance~\cite{Chiba:2003ir}) or have a scalar-tensor theory limit~\cite{Luty:2003vm,deRham:2010ik}. 
The focus of the present chapter regards exact{\footnote{We will focus on explicit solutions throughout this chapter. Numerical solutions will be treated elsewhere.}} BH solutions in four-dimensional scalar-tensor theories with a single scalar field $\phi$, with action $S\left[g_{\mu\nu},\phi\right]=\int\mathrm{d}^4x\sqrt{-g}\mathcal{L}$ (see~\cite{Herdeiro:2015waa} for previous reviews). Also, our attention will be restricted to asymptotically flat BHs in vacuum without a cosmological constant in the action{\footnote{Solutions with (a)dS asymptotics are in general less constrained than asymptotically flat solutions and generically easier to obtain: some scalar-tensor actions admit solutions with (effective or not) cosmological constant but do not allow asymptotically flat solutions, see e.g.~\cite{Rinaldi:2012vy}.}, and more precisely to BHs which behave asymptotically at leading order like the GR Schwarzschild/Kerr solutions, in order for them to display the expected Newtonian limit~\cite{Carroll:2004st}. We will use units in which $16\pi G=c=1$.
\\
The simplest theories are scalar-tensor Lagrangians which have at most two derivatives, however, as we will now see, no-hair theorems greatly restrict  the existence of non GR solutions for such Lagrangians. 
Since we are interested in modified gravity, we will consider a nonminimal coupling of the scalar to gravity\footnote{The case of minimally coupled scalar field will not be dealt with in this chapter. BH solutions can be constructed with such minimal coupling by reverse engineering the scalar field potential: one first imposes the solution, then finds the potential supporting it, see e.g.~\cite{Bechmann:1995sa}.}, and to fix this frame we shall suppose that any matter sources couple minimally to gravity. For nonminimal coupling and at most two derivatives, the theories are of a generalized Brans-Dicke form (see for example \cite{Sotiriou:2014yhm}),
\begin{equation}
\mathcal{L} = \phi R-\frac{\omega\left(\phi\right)}{\phi}\left(\partial\phi\right)^2-V\left(\phi\right).\label{eq:brans}
\end{equation}
where $R$ is the Ricci scalar of the metric $g_{\mu\nu}$ and $\omega\left(\phi\right)$ measures the strength of the scalar coupling, infinity representing the GR limit.
No-scalar-hair theorems exist for this kind of Lagrangians~\cite{Hawking:1972qk}. They greatly restrict the existing BH solutions to be GR solutions with a trivial scalar field. There is a particular Lagrangian with a non-minimally conformally coupled scalar $\phi$, and second-order in derivatives, which gives rise to an asymptotically flat BH. It is usually written as
\begin{equation}
\mathcal{L} = R-\frac{1}{2}\left(\partial\phi\right)^2-\frac{1}{12}R\phi^2\label{eq:bbmbac}
\end{equation}
(this is of the form~(\ref{eq:brans}) up to redefinition of the scalar field), and its static, spherically-symmetric solution is known as the Bocharova-Bronnikov-Melnikov-Bekenstein (BBMB) black hole~\cite{BBMB},
\begin{equation}
\mathrm{d}s^2=-f\mathrm{d}t^2+\frac{\mathrm{d}r^2}{f}+r^2\mathrm{d}\Omega^2,\quad f\left(r\right)=\left(1-\frac{M}{r}\right)^2,\quad \phi\left(r\right) = \frac{2\sqrt{3}M}{r-M},
\end{equation}
where $\mathrm{d}\Omega^2$ is the metric of the unit two-sphere\footnote{The generalization of the BBMB solution to (a)dS asymptotics is the Martinez-Troncoso-Zanelli (MTZ) BH~\cite{Martinez:2002ru}, and requires to include in the BBMB Lagrangian~(\ref{eq:bbmbac}) not only a cosmological constant $\Lambda$, but also a self-interacting potential $\lambda\phi^4$, with $\lambda=\Lambda/72$. The planar adS version requires the addition of axionic fields \cite{Bardoux:2012tr}. In the presence of a cosmological constant the scalar singularity can be hidden by the event horizon.}. This describes an extremal BH with double horizon at $r=M$ where $M$ is the mass of the BH. The scalar field diverges at the horizon, bringing about one of the pathologies of the BBMB solution.  

In a word, scalar-tensor Lagrangians which are second order in derivatives allow very few asymptotically flat BH solutions. A way to circumvent no-hair theorems is to consider higher-order  scalar-tensor theories. After all, if we are interested in UV corrections to gravity, such terms are bound to play an important role. This chapter reviews exact asymptotically flat BH solutions in four-dimensional, higher-order scalar-tensor theories. These theories are first described in section~\ref{sec:horn}. Section~\ref{sec:stealth} then presents the so-called stealth solutions, where a non-trivial scalar field supports a GR metric. On the other hand, non-stealth BHs, for which the metric differs from GR, are studied in section~\ref{sec:nonstealth}. Finally, a summary and discussions appear in the last section. 

\section{Horndeski and beyond}
\label{sec:horn}

\subsection{Horndeski theories}
\label{subsec:2}

As advocated in the introduction, higher-order scalar-tensor theories enable one to evade no-scalar-hair theorems and to obtain  BH solutions. In general, however, higher derivative Langrangians  lead to field equations with time derivatives of order higher than two, which lead (in general again) to the propagation of unphysical degrees of freedom with an associated Hamiltonian unbounded from below. This is the Ostrogradsky instability, and the unphysical degrees of freedom are called Ostrogradsky ghosts (see~\cite{Woodard:2015zca} and references therein). Of course, a sufficient condition for preventing the Ostrogradsky instability is for the field equations to be of second order. It turns out that the most general four-dimensional scalar-tensor action with second order field equations was described as early as in 1974 by Horndeski~\cite{Horndeski:1974wa}. His work went largely unnoticed until its rediscovery by~\cite{Charmousis:2011bf} in 2011. Fortunately, independent progress on scalar-tensor theories was accomplished during those almost forty years following differing paths and arriving to the same conclusions. Starting from the Galileon~\cite{Nicolis:2008in} (most general scalar action in flat spacetime with invariance under \textit{Galilean boost} $\phi\to \phi+b^\mu x_\mu+c$ with constant $b^\mu$ and $c$), 'covariantization'~\cite{Deffayet:2009wt} and several steps led to the generalized Galileon~\cite{Deffayet:2011gz}, whose action is simply written in terms of four arbitrary functions $G_2$, $G_3$, $G_4$, $G_5$. It was later shown~\cite{Kobayashi:2011nu} that this simple generalized Galileon action is equivalent to the Horndeski action. As a consequence, the Horndeski action is now conventionally written in its generalized Galileon form, namely,
\begin{align}
S{}&{}\left[g_{\mu\nu},\phi\right] = \int \mathrm{d}^4x\sqrt{-g}\, \Big\lbrace \mathcal{L}_2+\mathcal{L}_3+\mathcal{L}_4+\mathcal{L}_5\Big\rbrace,\label{eq:horndeski_lag}\\
\mathcal{L}_2 ={}&{} G_2, \quad \mathcal{L}_3=-G_3\Box\phi,\quad \mathcal{L}_4 =  G_4R+G_{4X}\left[\left(\Box\phi\right)^2 -\left(\phi_{\mu\nu}\right)^2 \right], \\
\mathcal{L}_5 ={}&{} G_5\,G_{\mu\nu}\phi^{\mu\nu} -
\frac{1}{6}G_{5X}\left(\left(\Box\phi\right)^3 -
3\Box\phi\left(\phi_{\mu\nu}\right)^2
+2\phi_{\mu\nu}\phi^{\nu\rho}\phi{^\mu_\rho}\right),
\end{align}
where $G_2$, $G_3$, $G_4$, $G_5$ are arbitrary functions of the scalar field $\phi$ and its kinetic term $X\equiv -\left(\partial\phi\right)^2/2$: $G_2=G_2\left(\phi,X\right)$, etc. A subscript $X$ means derivation with respect to $X$. $G_{\mu\nu}$ is the Einstein tensor of the metric $g_{\mu\nu}$, and the following notations are used for brevity: $\phi_\mu=\partial_\mu\phi$, $\phi_{\mu\nu}=\nabla_\mu\nabla_\nu\phi$, $\left(\phi_{\mu\nu}\right)^2 = \phi_{\mu\nu}\phi^{\mu\nu}$. Note that, if all functions $G_2$, $G_3$, $G_4$, $G_5$ depend only on the kinetic term $X$, the Horndeski action is shift-symmetric, i.e. unchanged under shifts of the scalar $\phi\to\phi+\text{const}$. This continuous symmetry brings about a conserved Noether current, 
\begin{equation}
J^\mu = -\frac{1}{\sqrt{-g}}\frac{\delta S}{\delta \left(\partial_\mu\phi\right)},\label{eq:noether}
\end{equation}
and the scalar field equation becomes $0=\nabla_\mu J^\mu$. As a result the shift-symmetric version of Horndeski is accordingly easier to deal with. In addition, if $G_3=G_5=0$, a shift-symmetric Horndeski action acquires the discrete parity symmetry $\phi\to -\phi$, which turns out to simplify things as well. Last but not least, by definition of Horndeski theory, every scalar-tensor action whose field equations are second order can be cast into the form~(\ref{eq:horndeski_lag}), modulo a number of (obvious or less obvious) integrations by parts. For instance, the derivative coupling with the Einstein tensor, $G^{\mu\nu}\partial_\mu\phi\,\partial_\nu\phi$, corresponds to a galieon $G_4 = X$, while a coupling $\phi\mathcal{G}$ with the Gauss-Bonnet invariant $\mathcal{G}\equiv R^2-4R_{\mu\nu}R^{\mu\nu}+R_{\mu\nu\rho\sigma}R^{\mu\nu\rho\sigma}$ coincides with $G_5=-4\log\left\lvert X\right\rvert$, see~\cite{Kobayashi:2011nu,Langlois:2022eta}. More details on Horndeski theories can be found in the review~\cite{Kobayashi:2019hrl}.

\subsection{Degenerate Higher-Order Scalar-Tensor (DHOST) theories}
Second order field equations are a sufficient condition for avoiding Ostrogradsky instability, but are by no means necessary, at least for scalar-tensor theories{\footnote{Going beyond scalar-tensor theories (vector/tensor, bigravity theories, etc.), it turns out that avoiding instabilities is far more complex~\cite{Crisostomi:2017ugk}.}. Indeed, with field equations whose time derivatives are of order higher than two, the Ostrogradsky ghost emerges as an extra degree of freedom associated with the need of additional initial conditions. However, degeneracy of the kinetic matrix of the system eliminates this extra degree of freedom, yielding theories without Ostrogradsky instability despite having field equations of order higher than two. Exploiting this fact in the context of scalar-tensor theories leads to the Degenerate Higher-Order Scalar-Tensor (DHOST) theories (see~\cite{Langlois:2018dxi} for a review), which evidently comprise the Horndeski theory. They were first constructed for Lagrangians quadratic in the second derivatives~\cite{Langlois:2015cwa}, then extended to cubic order~\cite{BenAchour:2016fzp}. We will not need the full expression of the DHOST Lagrangian, and will rather write it down partly when needed. From our point of view which focuses on solutions rather than theories themselves, it is more interesting to note that DHOST theories admit a quite simple interpretation in terms of \textit{conformal-disformal transformations} of Horndeski theory~\cite{BenAchour:2016cay}. A conformal-disformal transformation~\cite{Bekenstein:1992pj} depends on two functions of the scalar and its kinetic term, the conformal factor $C\left(\phi,X\right)$ and the disformal factor $D\left(\phi,X\right)$, mapping a metric $g_{\mu\nu}$ to a new metric $\widetilde{g}_{\mu\nu}$ according to
\begin{equation}
\widetilde{g}_{\mu\nu} = C\left(\phi,X\right)g_{\mu\nu}+D\left(\phi,X\right)\partial_\mu\phi\,\partial_\nu\phi.\label{eq:disf}
\end{equation}
Given an initial Horndeski action $S\left[g_{\mu\nu},\phi\right]$, it is possible to express it in terms of the metric $\widetilde{g}_{\mu\nu}$,
\begin{equation}
S\left[g_{\mu\nu},\phi\right] = \widetilde{S}\left[\widetilde{g}_{\mu\nu},\phi\right],\label{eq:new_ac}
\end{equation}
and it turns out that the resulting action $\widetilde{S}$ is a DHOST theory. More precisely, three cases can be distinguished. If the two factors depend only on the scalar field, $C=C\left(\phi\right)$ and $D=D\left(\phi\right)$, then the new action also belongs to the Horndeski class~\cite{Bettoni:2013diz}, (\ref{eq:disf}) is in this case an internal map. If $C=C\left(\phi\right)$, but the disformal factor depends on $X$, $D=D\left(\phi,X\right)$, then the new action belongs to the so-called beyond Horndeski class~\cite{Gleyzes:2014dya}, which, in addition to the four Horndeski terms~(\ref{eq:horndeski_lag}), admit two additional contributions,
\begin{align}
S_\text{bH} ={}&{} \int\mathrm{d}^4x\sqrt{-g}\Big\lbrace \mathcal{L}_{4\text{bH}}+\mathcal{L}_{5\text{bH}}\Big\rbrace,\label{eq:bh}\\
\mathcal{L}_{4\text{bH}} ={}&{}F_4\,\varepsilon^{\mu\nu\rho\sigma}\,\varepsilon^{\alpha\beta\gamma}_{\,\,\,\,\,\,\,\,\,\,\,\sigma}\,\phi_\mu\phi_\alpha\phi_{\nu\beta}\phi_{\rho\gamma},\\
\mathcal{L}_{5\text{bH}}={}&{}F_5\,\varepsilon^{\mu\nu\rho\sigma}\,\varepsilon^{\alpha\beta\gamma\delta}\phi_\mu\phi_\alpha\phi_{\nu\beta}\phi_{\rho\gamma}\phi_{\sigma\delta},
\end{align}
with functions\footnote{Considering the full beyond Horndeski Lagrangian, i.e. the pure Horndeski contribution~(\ref{eq:horndeski_lag}) and the pure beyond Horndeski part~(\ref{eq:bh}), the degeneracy of the kinetic matrix imposes the relation $XG_{5X}F_4=3F_5\left(G_4-2XG_{4X}\right)$~\cite{Crisostomi:2016tcp} leaving us at the end with one additional degree of freedom equivalent to the disformal function $D$.} $F_4=F_4\left(\phi,X\right)$ and $F_5=F_5\left(\phi,X\right)$, and $\varepsilon_{\mu\nu\rho\sigma}$ is the Levi-Civita tensor. Finally, if $C=C\left(\phi,X\right)$, the resulting action is a DHOST action, but not of the beyond Horndeski type~\cite{BenAchour:2016cay}. 

More generally, DHOST theories are stable under conformal-disformal transformations: if the initial action $S\left[g_{\mu\nu},\phi\right]$ is a DHOST action, then after the conformal-disformal transformation~(\ref{eq:disf}), the resulting action $\widetilde{S}\left[\widetilde{g}_{\mu\nu},\phi\right]$ given by~(\ref{eq:new_ac}) also belongs to the DHOST family~\cite{BenAchour:2016cay}, or in other words, degeneracy of the kinetic matrix remains unaffected by a conformal-disformal transformation. This stability enables \textit{generation of solutions}~\cite{BenAchour:2020wiw} in the following way: starting from a \textit{seed solution} $\left(g_{\mu\nu},\phi\right)$ to the initial action $S$, one gets a solution $\left(\widetilde{g}_{\mu\nu},\phi\right)$ to the new action $\widetilde{S}$ since
\begin{equation}
\widetilde{\mathcal{E}}_{\mu\nu} = \frac{\sqrt{-g}}{\sqrt{-\widetilde{g}}}\frac{\delta g^{\rho\sigma}}{\delta \widetilde{g}^{\mu\nu}}\mathcal{E}_{\rho\sigma},\quad \widetilde{\mathcal{E}}_\phi = \frac{\sqrt{-g}}{\sqrt{-\widetilde{g}}}\mathcal{E}_\phi,\label{eq:disf_eq}
\end{equation}
where $\mathcal{E}_{\mu\nu}=\left(-g\right)^{-1/2}\delta S/\delta g^{\mu\nu}$, etc., are the field equations. This useful fact will be exploited in the next section.

\section{Stealth black holes and their disformal and conformal transformations}\label{sec:stealth}
As we saw, no-hair theorems establish that, under certain assumptions, a scalar-tensor theory admit no other BH solution than a GR metric (Schwarzschild or Kerr) with a trivial scalar field. Hence, finding a GR BH dressed with a non-trivial scalar field already represents a step forward. Writing the field equations as
\begin{equation}
G_{\mu\nu}=T_{\mu\nu}
\end{equation} 
where $T_{\mu\nu}$ is the energy-momentum tensor associated to the scalar field, such solutions require $T_{\mu\nu}=0$, so that the scalar field, although non-trivial, does not modify the GR field equations{\footnote{One can slightly modify the above relation to allow for de Sitter stealth solutions, i.e. Einstein metric solutions with a non trivial scalar field. Such solutions have self tuning properties for the cosmological constant, see for example \cite{Charmousis:2015aya,Babichev:2016kdt}.}}. These solutions are therefore known as \textit{stealth solutions}, as opposed to non-stealth solutions, whose metric is not a GR metric and which will be studied in later sections. For stealth solutions, the non-trivial scalar field still enters the perturbation theory and may therefore modify, for instance, the emission of GW. Furthermore, one may, using stealth solutions, construct non-stealth images by using conformal/disformal transformations. This latter possibility may be very useful as the field equations for stationary black holes are extremely complex for analytic study. The current section first presents Schwarzschild and Kerr stealth solutions, before moving on to the generation of a \textit{disformed Kerr metric} by a disformal transformation, as well as conformal cosmological versions.

\subsection{Generic stealth Schwarzschild in Horndeski theories}

The first asymptotically flat BH in higher-order scalar-tensor theories was constructed in~\cite{Babichev:2013cya}. The action is
\begin{equation}
S = \int\mathrm{d}^4x\sqrt{-g}\Bigl\{ R+\beta G^{\mu\nu}\partial_\mu\phi\,\partial_\nu\phi\Bigr\},
\end{equation}
which belongs to the Horndeski class with a unique nonvanishing function, $G_4=1+\beta X$, with $\beta$ a coupling. The action is shift-symmetric, and the authors are looking for a static, spherically-symmetric solution,
\begin{equation}
\mathrm{d}s^2=-h\left(r\right)\mathrm{d}t^2+\frac{\mathrm{d}r^2}{f\left(r\right)}+r^2\mathrm{d}\Omega^2.
\end{equation}
Under this setting, there exists a no-hair argument~\cite{Hui:2012qt}, which requires the additional assumption that $\phi=\phi\left(r\right)$. However, following~\cite{Babichev:2010kj}, the authors of~\cite{Babichev:2013cya} realized that such a radial profile for the scalar field is not the most general compatible with the symmetries of the spacetime. Indeed, the action depends on the scalar field only via its derivatives, so do the field equations, therefore, a linear time dependence of the scalar is allowed,
\begin{equation}
\phi = qt+\psi\left(r\right),\label{eq:lin_time}
\end{equation}
with $q$ a constant. This form of the scalar is then found to support a stealth Schwarzschild solution,
\begin{equation}
h\left(r\right)=f\left(r\right)=1-\frac{2M}{r},
\end{equation}
with the usual mass integration constant $M$, and the radial part of the scalar field reads
\begin{equation}
\psi'\left(r\right) = \pm q\frac{\sqrt{2Mr}}{r-2M}.\label{eq:psi_st_sch}
\end{equation}
The apparent logarithmic divergence of $\psi\left(r\right)$ at the Schwarzschild radius $r=2M$ stems from the poor behaviour of the coordinates $\left(t,r,\theta,\varphi\right)$ at the horizon. Changing for horizon-crossing coordinates~\cite{Carroll:2004st}, $\left(v,r,\theta,\varphi\right)$ or $\left(u,r,\theta,\varphi\right)$, where $v=t+r^\star$ and $u=t-r^\star$ are the usual advanced and retarded time in terms of the tortoise coordinate $r^\star=\int\mathrm{d}r/\sqrt{h\left(r\right)f\left(r\right)}$, the Schwarzschild metric is well-behaved and the scalar field is regular at the future horizon, $\phi\left(v,r=2M\right)=qv+\text{const.}$, if the $+$ sign is chosen in~(\ref{eq:psi_st_sch}), or at the past horizon, $\phi\left(u,r=2M\right)=qu+\text{const.}$, if the $-$ sign is chosen in~(\ref{eq:psi_st_sch}). Note that the kinetic term of the solution is constant,
\begin{equation}
X = -\frac{1}{2}\left(\partial\phi\right)^2 = \frac{1}{2}\left(\frac{q^2}{h\left(r\right)}-f\left(r\right)\psi'\left(r\right)^2\right) = \frac{q^2}{2}.
\end{equation}
This property was exploited in~\cite{Kobayashi:2014eva} to construct other solutions, or more precisely, to find more general forms of function $G_4$ supporting the existence of the stealth Schwarzschild solution\footnote{Generalizations to stealth Schwarzschild-de-Sitter are presented in both~\cite{Babichev:2013cya} and~\cite{Kobayashi:2014eva}, in which case a bare cosmological constant $\Lambda$ is added into the action, as well as a nontrivial kinetic term. The effective cosmological constant of the BH is then a combination of the bare one $\Lambda$, and of the $G_2$ and $G_4$ terms as evaluated at the constant value of $X$. The solutions have self tuning properties~\cite{Charmousis:2015aya,Babichev:2016kdt}.}. Stealth solutions prove to be a good starting point for our solution quest, however, they have at least two shortcomings : first they are too much GR-like by definition, and second their construction imposes the absence of the scalar field kinetic term in the action which brings about perturbative problems. The former problem was shown to lead to non hyperbolic scalar field perturbations~\cite{Babichev:2018uiw} akin to strong coupling. This is unlike the de Sitter stealth versions, which do include a kinetic term~\cite{Babichev:2017lmw}. Finally, in~\cite{Babichev:2016kdt} stealth solutions were presented in beyond Horndeski theories, while Refs.~\cite{Minamitsuji:2018vuw} and~\cite{Motohashi:2019sen} found conditions for the existence of stealth Schwarzschild solutions in DHOST theories, Refs.~\cite{Minamitsuji:2018vuw} studying in particular the non-shift symmetric case.

\subsection{A stealth Kerr black hole in a GR like DHOST theory}
The authors of~\cite{Charmousis:2019vnf} found a way of generalising the previous construction to a stealth Kerr BH in DHOST theory. Note that this construction is not possible in Horndeski theory; in fact it turns out that a stealth stationary metric is a lot harder to obtain than a stealth static metric which appears quite generically in shift and parity symmetric Horndeski theories. We now use the usual notations of DHOST theory~\cite{Langlois:2015cwa}, with in particular a kinetic term $X\equiv\left(\partial\phi\right)^2$, which also turns out to be more clear for the following. Ref.~\cite{Charmousis:2019vnf} considers the most general shift-symmetric quadratic DHOST theory with speed of GW equal to the speed of light~\cite{Langlois:2017dyl},
\begin{align}
S=\int\mathrm{d}^4x \Bigl\{f\left(X\right)R+{}&{}P\left(X\right)+Q\left(X\right)\Box\phi+A_3\left(X\right)\phi^\mu\phi_{\mu\nu}\phi^\nu\Box\phi\nonumber\\{}&{}+A_4\left(X\right)\phi^\mu\phi_{\mu\nu}\phi^{\nu\rho}\phi_\rho+A_5\left(X\right)\left(\phi^\mu\phi_{\mu\nu}\phi^{\nu}\right)^2\Bigr\},
\end{align}
with four free functions $f$, $P$, $Q$, $A_3$, while degeneracy of the kinetic matrix yields $A_4$ and $A_5$~\cite{Langlois:2015cwa},
\begin{align}
A_4 ={}&{} \frac{1}{8f}\left(48f_X^2-8\left(f-Xf_X\right)A_3-X^2A_3^2\right),\\ A_5 ={}&{} \frac{A_3}{2f}\left(4f_X+XA_3\right).
\end{align}
It turns out that only within this theory, which has graviton speed identical to GR, it is seemingly possible to obtain a stealth Kerr solution.
Taking inspiration from the stealth Schwarzschild solution, the authors of~\cite{Charmousis:2019vnf} are looking for a Kerr solution with constant kinetic term $X=X_0\equiv-q^2$. This greatly simplifies the DHOST field equations, since the $A_4$ and $A_5$ terms do not contribute for constant $X$, and the Kerr metric implies a vanishing Ricci tensor. It is then found that such a stealth Kerr solution is allowed provided the DHOST functions verify\footnote{Ref.~\cite{Charmousis:2019vnf} constructs more generally a stealth Kerr-dS solution, in which case conditions~(\ref{eq:conditions}) are replaced by analogous ones involving the cosmological constant $\Lambda$.}
\begin{equation}
P\left(X_0\right)=P_X\left(X_0\right)=Q_X\left(X_0\right)=A_3\left(X_0\right) = 0. \label{eq:conditions}
\end{equation}
It thus remains to solve the requirement of constant kinetic term, $X=X_0=-q^2$ for the Kerr geometry. This can be linked to the mass-shell equation of a point-particle with mass $q$,
\begin{equation}
-q^2 = \left(\partial\phi\right)^2 = g^{\mu\nu}\partial_\mu\phi\,\partial_\nu\phi\equiv g^{\mu\nu}p_\mu p_\nu,
\end{equation}
where $p^\mu$ is the four-momentum of the point-particle, $p^\mu = \mathrm{d}x^\mu/\mathrm{d}\lambda$, with $\lambda$ an affine parameter along the geodesic followed by the point-particle. Accordingly, $\phi$ is a \textit{Hamilton-Jacobi functional} for the geodesic trajectory, i.e., its exterior derivative is the momentum one-form of the geodesic, $\mathrm{d}\phi=\partial_\mu\phi\,\mathrm{d}x^\mu=p_\mu\mathrm{d}x^\mu$. It is then only a matter of using Carter's integration of Kerr geodesics~\cite{Carter:1968rr} to find the form of $p_\mu\mathrm{d}x^\mu$ and therefore of $\phi$. The final form of the regular scalar field, with $X=-q^2$, is\footnote{In the stealth Kerr-dS case, regularity of the scalar field at both the event and the cosmological horizons requires it to depend also on the colatitude angle $\theta$, $\phi=\phi\left(t,r,\theta\right)$.}
\begin{equation}
\phi = q\left(t\pm\int\frac{\sqrt{2Mr\left(r^2+a^2\right)}}{\Delta}\mathrm{d}r\right),\label{eq:hj}
\end{equation}
where as usual, $M$ and $a$ are the mass and angular momentum per unit mass of the Kerr spacetime, and $\Delta = r^2+a^2-2Mr$. All this is written in the usual Boyer-Lindquist coordinates $\left(t,r,\theta,\varphi\right)$, where the Kerr metric is~\cite{Carroll:2004st}
\begin{align}
\mathrm{d}s^2={}&{}-\left(1-\frac{2Mr}{\Sigma}\right)\mathrm{d}t^2+\frac{\Sigma}{\Delta}\mathrm{d}r^2+\Sigma\mathrm{d}\theta^2-\frac{4Mar\sin^2\theta}{\Sigma}\mathrm{d}t\mathrm{d}\varphi\nonumber\\{}&{}+\frac{\sin^2\theta}{\Sigma}\left[\left(r^2+a^2\right)^2-a^2\Delta\sin^2\theta\right]\mathrm{d}\varphi^2,\label{eq:kerr}
\end{align}
with $\Sigma = r^2+a^2\cos^2\theta$. The expression~(\ref{eq:hj}) reduces to the stealth Schwarzschild scalar~(\ref{eq:lin_time})-(\ref{eq:psi_st_sch}) for vanishing rotation, and just as in this case, regularity at the horizons $\Delta=0$ can be verified by changing to horizon-crossing coordinates for the Kerr metric. Other interesting works include Ref.~\cite{Takahashi:2020hso}, which undertook a systematic study of the most general quadratic DHOST theory in order to determine conditions for it to admit the stealth Kerr solution\footnote{And, more denerally, the Kerr-Newman-dS solution.}. The linear metric perturbations of the stealth Kerr solution were studied in~\cite{Charmousis:2019fre} and they were found to have a modified Teukolsky form. Regarding the stability of stealth solutions in DHOST theories, odd-parity perturbations around static stealth solutions were discussed in~\cite{Takahashi:2019oxz}, while Refs.~\cite{deRham:2019gha} and~\cite{Takahashi:2021bml} extended the analysis to even-parity perturbations, stating that they were strongly coupled. In fact, this strong coupling problem also exists for the stealth Kerr solution~\cite{deRham:2019gha} and is most probably due to the fact that the kinetic term is absent for this solution which entails the non hyperbolic character of the scalar perturbation equation. Discussions on the significance of strong coupling in stealth solutions and ways to circumvent the problem have been recently discussed in \cite{DeFelice:2022xvq}.

\subsection{Disformal Kerr metric}\label{subsec:disf}
As we pointed out earlier, see discussion around Eq.~(\ref{eq:disf_eq}), a conformal-disformal transformation~(\ref{eq:disf}) maps a seed solution $\left(g_{\mu\nu},\phi\right)$ of an initial DHOST theory to another solution $\left(\widetilde{g}_{\mu\nu},\phi\right)$ for a new DHOST theory. Let us first consider the case of a purely disformal transformation with, $C\left(\phi,X\right)=1$. For a start a disformal transformation depending solely on $X$, of a stealth Schwarzschild  solution turns out to be a coordinate transformed stealth Schwarzschild  solution with a rescaled mass parameter \cite{Babichev:2017lmw}. Therefore a disformal map in this case simply carries the same Ricci flat geometry from Horndeski to beyond Horndeski theory (essentially because $X$ is a constant). The disformal transformation of the seed stealth Kerr solution turns out to be far more interesting and was undertaken in~\cite{Anson:2020trg}, \cite{BenAchour:2020fgy} with a constant disformal factor $D\left(X\right)=\text{const.}$, since again $X$ is constant. Note in passing that a dependence of $D$ on the scalar field~(\ref{eq:hj}) would introduce an explicit time-dependence in the metric and thus spoil the spacetime symmetries. We will come back to that in the next section. The disformed Kerr metric is written as
\begin{equation}
\widetilde{g}_{\mu\nu}=g_{\mu\nu}-\frac{D}{q^2}\partial_\mu\phi\,\partial_\nu\phi,\label{eq:disf_kerr_impl}
\end{equation}
with $D$ a dimensionless constant, and $g_{\mu\nu}$ is the Kerr metric~(\ref{eq:kerr}) with mass $M$ and angular momentum per unit mass $a$. The DHOST theory admitting the disformal Kerr metric as a solution can be found in~\cite{BenAchour:2020fgy}. Its black hole properties were analysed in~\cite{Anson:2020trg} which we now turn to. Eq.~(\ref{eq:disf_kerr_impl}) produces the following line element,
\begin{align}
\mathrm{d}\widetilde{s}^2={}&{}-\left(1-\frac{2\widetilde{M}r}{\Sigma}\right)\mathrm{d}t^2-\frac{4\sqrt{1+D}\widetilde{M}ar\sin^2\theta}{\Sigma}\mathrm{d}t\mathrm{d}\varphi+\Sigma\mathrm{d}\theta^2\nonumber\\{}&{}+\frac{\sin^2\theta}{\Sigma}\left[\left(r^2+a^2\right)^2-a^2\Delta\sin^2\theta\right]\mathrm{d}\varphi^2-2D\frac{\sqrt{2\widetilde{M}r\left(r^2+a^2\right)}}{\Delta}\mathrm{d}t\mathrm{d}r\nonumber\\{}&{}+\frac{\Sigma\Delta-2D\left(1+D\right)\widetilde{M}r\left(r^2+a^2\right)}{\Delta^2}\mathrm{d}r^2,\label{eq:disf_kerr}
\end{align}
with $\widetilde{M}=M/\left(1+D\right)$, while $\Sigma$ and $\Delta$ are as in the seed Kerr metric, in particular $\Delta$ still features the parameter $M$ and not $\widetilde{M}$. $\Sigma=0$ corresponds to the ring singularity, just like for Kerr~\cite{Carroll:2004st}. Regarding the conserved charges, the disformed Kerr metric has mass $\widetilde{M}$, while its angular momentum per unit mass is $\widetilde{a}\equiv a\sqrt{1+D}$. The disformal transformation thus rescales the mass and angular momentum, $J = aM\to\tilde{J} = J/\sqrt{1+D}$. In fact, for vanishing rotation, $a=0$, that is to say, if the seed metric is stealth Schwarzschild, then a change of coordinates removes the $\mathrm{d}t\mathrm{d}r$ term of~(\ref{eq:disf_kerr}) and the disformed metric is again Schwarzschild, with a mere rescaling of the mass, see also~\cite{Babichev:2017lmw} and~\cite{BenAchour:2020wiw}. The disformal transformation turns out to be non-trivial only if $D$ and $a$ are both non zero as (\ref{eq:disf_kerr}) is a non Ricci flat metric. Indeed, if~(\ref{eq:disf_kerr}) were the Kerr metric in different coordinates, then it would also share its property of circularity, which for the Kerr metric manifests itself by the invariance under $\left(t,\varphi\right)\to\left(-t,-\varphi\right)$. The disformed Kerr metric as written in~(\ref{eq:disf_kerr}) does not possess this invariance because of the $\mathrm{d}t\mathrm{d}r$ term, and the authors of~\cite{Anson:2020trg} prove that this is not due to the choice of coordinates, by referring to the more fundamental definition of circularity, which means integrability of the 2-submanifold orthogonal to the two Killing vector fields $\partial_t$ and $\partial_\varphi$ of an axisymmetric spacetime.

Non-circular spacetimes have a richer causal structure than circular ones~\cite{Johannsen:2013rqa}. For disformed Kerr just as for Kerr, there is an ergosphere or static limit, where $\partial_t$ is null, and inside\footnote{'Inside' here means 'for decreasing $r$'.} which static observers, i.e. observers with constant $r$, $\theta$ and $\varphi$, cease to exist. Inside the ergosphere, the stationary limit is defined as the hypersurface inside which stationary observers, that is, with constant $r$ and $\theta$, cease to exist. Equivalently, this stationary limit is a Killing horizon: there exists a linear combination $\partial_t+\Omega\partial_\varphi$ which is null, with $\Omega$ then defined as the angular velocity of the BH. For circular spacetimes like Kerr, this Killing horizon coincides with the event horizon (this is a particular case of the rigidity theorem, see e.g.~\cite{Hawking:1973uf}), however, when the spacetime is non-circular, the stationary limit is no more the event horizon, which must be looked for as a null hypersurface inside the Killing horizon. 

Concerning the particular case of the disformed Kerr metric~\cite{Anson:2020trg}, the static limit or ergosphere lies at
\begin{equation}
r\left(\theta\right) = \widetilde{M}+\sqrt{\widetilde{M}^2-a^2\cos^2\theta},
\end{equation}
just as for Kerr up to the rescaling of the mass, while the stationary limit or Killing horizon is located at $r=R_0\left(\theta\right)$ such that
\begin{equation}
P\left(R_0\left(\theta\right),\theta\right)=0
\end{equation}
where 
\begin{equation}
P\left(r,\theta\right)\equiv r^2+a^2-2\widetilde{M}r+\frac{2D\widetilde{M}ra^2\sin^2\theta}{\Sigma},
\end{equation}
yielding a fourth-order algebraic equation for $R_0\left(\theta\right)$. Finally, the event horizon $r=R\left(\theta\right)$ is given by a nonlinear ordinary differential equation,
\begin{equation}
\left(\frac{\mathrm{d}R\left(\theta\right)}{\mathrm{d}\theta}\right)^2+P\left(R\left(\theta\right),\theta\right)=0.
\end{equation}
Ref.~\cite{Anson:2020trg} investigated numerically the parameter space allowing solution to this equation, finding in particular that for a nonvanishing disformal parameter $D$, this candidate event horizon exists only for an angular momentum per unit mass $\widetilde{a}<\widetilde{a}_c<1$, with the upper bound depending on the value of $D$, $\widetilde{a}_c = \widetilde{a}_c\left(D\right)$. Of course, the parameter $D$ could be constrained by observations, since it deforms the shadow with respect to the one of a Kerr BH~\cite{Long:2020wqj}, or modifies the orbit of stars around the BH~\cite{Anson:2021yli}.

\subsection{Conformal Kerr metric: cosmological black holes}\label{subsec:conf}
In the previous subsection a purely disformal transformation was applied to obtain new stationary solutions distinct from the Kerr metric. 
There we took a trivial conformal factor $C=1$ in order to retain asymptotically flat solutions. 
In this section we instead assume a non-trivial conformal factor depending explicitly on the scalar field,  $C=C(\phi)$, while we set the disformal factor to zero, $D=0$.
The causal structure is preserved under such a transformation, i.e. starting with the seed solution of a theory with the speed of gravity equals unity, one obtains a theory with the same property. 
Having in mind an asymptotically flat seed metric with a non-trivial asymptotic dependence of the scalar field on time, such a method naturally gives rise to a time-dependent black hole metric. In particular, rotating growing black holes embedded in Friedmann-Lema\^{i}tre-Robertson-Walker (FLRW) universe can be constructed~\cite{Babichev:2023mgk}. 
Indeed, if the gradient of the scalar field $\phi$ is time-like,
one may choose $\phi$ to be (proportional to) a time coordinate, so that the new metric $C(\phi)g_{\mu\nu}$ asymptotically has a form $C(\phi)g_{\mu\nu}^\text{(flat)}$ with $\phi$ being the conformal time, i.e. FLRW metric in conformal time is recovered asymptotically. 

A suitable case  of the seed solution is a stealth Kerr black hole, which has been already described in detail above. It has the correct flat metric asymptotic with a non-trivial time-dependent scalar field. 
The metric that is obtained by conformal transformation is neither spherically symmetric nor stationary. 
To uncover the properties of such a conformal metric one defines the trapping (or apparent) horizons, which use the notion of 'expansion', related to the definition of null geodesic congruences~\cite{Faraoni:2015ula}. 
While in the general case of non-zero rotation it is quite a non-trivial task to find the trapping horizons, the problem is greatly simplified when the rotation is set to zero.
In this case, the metric for the resulting conformal solution has the Schwarzschild form, written in Painlev\'e-Gullstrand coordinates, multiplied by the conformal factor that depends on the Painlev\'e-Gullstrand time:
\begin{align}
\mathrm{d}\widetilde{s}^2 ={}& A^2(\tau)\Biggl\{-\left(1-\frac{2M}{r}\right)\mathrm{d}\tau^2+2\sqrt{\frac{2M}{r}}\mathrm{d}\tau \mathrm{d}r+\mathrm{d}r^2+r^2\mathrm{d}\Omega^2\Biggr\}.\label{eq:sch}
\end{align}
Clearly, at large $r$ the metric becomes the FLRW line element in conformal time $\tau$. 
The scale factor of the universe $A(\tau)$ is related to the choice of the conformal factor $C(\phi)$ and is defined by the DHOST theory, for which the conformal solution is written.
The obtained metric is non-singular everywhere apart from the Big Bang singularity $A=0$ and the curvature black hole singularity $r=0$. 
The metric~(\ref{eq:sch}) has been discussed in the context of GR as an \emph{ad-hoc} line element in~\cite{Culetu:2012ih}, while 
its properties were recently studied~\cite{Sato:2022yto} for particular choices of the scale factor $A(\tau)$.

The analysis of the conformal spacetime~(\ref{eq:sch}) reveals two trapping horizons,
\begin{equation}
\tau_+ = -\alpha r\left(1-\sqrt{\frac{2M}{r}}\right)^{-1},\quad \tau_- = \alpha r \left(1+\sqrt{\frac{2M}{r}}\right)^{-1},\label{eq:taupm}
\end{equation}
where the '+' sign corresponds to the black hole horizon and '-' sign corresponds to the cosmological horizon, while $\alpha$ is the exponent in the scale factor, $A\left(\tau\right)\propto \left\lvert\tau\right\rvert^\alpha$.
One should note that $r$ is not a physical distance but a comoving coordinate, therefore a relevant measure for the physical radius is 
$R_{phys}=A\left(\tau\right)r$. 
The case of non-zero rotation can be treated similarly, however because of the absence of spherical symmetry the analysis is considerably more complicated, see the details in~\cite{Babichev:2023mgk}.
As a further generalisation of the method presented above, the stealth Kerr-de-Sitter solution of a DHOST theory can be considered as a seed metric. 
In this case, the scale factor of the asymptotically FLRW line element is composed from two parts, the original de Sitter expansion and the extra expansion due to the conformal factor $C(\phi)$, see details in Ref.~\cite{Babichev:2023mgk}.

\section{Black holes beyond Ricci flat metrics}\label{sec:nonstealth}
Stealth BHs turn out to be a good starting point and interesting as seed solutions, nevertheless their Ricci flat geometry makes them very much alike GR albeit the problems associated with the scalar perturbations. It is therefore paramount to move on and try to seek solutions which differ from the Schwarzschild/Kerr metrics predicted by GR\footnote{Note that every asymptotically flat BH presented from now on admits a direct generalization to (a)dS asymptotics by simply introducing a cosmological constant $\Lambda$ into the action.}. 

\subsection{Parity and shift-symmetric theories}\label{subsec:parity}
As explained when introducing the Horndeski action~(\ref{eq:horndeski_lag}), the field equations simplify under the assumption that all Horndeski functions $G_2$, $G_3$, $G_4$, $G_5$, and also the beyond Horndeski functions $F_4$ and $F_5$, depend only on the kinetic term $X$ and not on the scalar field $\phi$, because this implies invariance of the action under shifts $\phi\to\phi+\text{const.}$, and therefore a conserved Noether current $J^\mu$, see Eq.~(\ref{eq:noether}). In addition, the terms $G_3$, $G_5$ and $F_5$ have long resisted the exact resolution of field equations: before 2020, no exact asymptotically flat solution with Newtonian asymptotics could be found when including those terms. We thus begin with a presentation of the non-stealth solutions for the following subclass of beyond Horndeski theories,
\begin{align}
S = \int\mathrm{d}^4 x\sqrt{-g}\Bigl\{G_2\left(X\right)+G_4{}&{}\left(X\right)R+G_{4X}\left[\left(\Box\phi\right)^2-\phi_{\mu\nu}\phi^{\mu\nu}\right]\nonumber\\{}&{}+F_4\left(X\right)\epsilon^{\mu\nu\rho\sigma}\epsilon^{\alpha\beta\gamma}_{\hspace{0.5cm}\sigma}\phi_\mu\phi_\alpha\phi_{\nu\beta}\phi_{\rho\gamma}\Bigr\}\label{eq:shift-parity}
\end{align}
which is not only shift-symmetric but also parity-symmetric, $\phi\to-\phi$, thanks to the absence of $G_3$, $G_5$ and $F_5$. It is important to note for the exact solutions presented here that scalar quantities associated to $\phi$, like $X$ or the norm $J_\mu J^\mu$ of the Noether current, are finite everywhere for $r\geq r_h$, where $r_h$ is the horizon radius. This can be important concerning the regularity of the solutions as advocated in \cite{Hui:2012qt}. In fact, all solutions presented here have $J_\mu J^\mu=0$ everywhere, in other words $J^r=0$ and as such have one less integration constant associated to the scalar field.  Moreover, they are all static and spherically-symmetric,
\begin{equation}
\mathrm{d}s^2=-h\left(r\right)\mathrm{d}t^2+\frac{\mathrm{d}r^2}{f\left(r\right)}+r^2\mathrm{d}\Omega^2.\label{eq:ansatz}
\end{equation}
and this is also true for $\phi=\phi(r)$ although one could take a linear time dependence for the scalar.

\subsubsection{Pure Horndeski black hole}
The first such asymptotically flat BH solution to action~(\ref{eq:shift-parity}) was found in~\cite{Babichev:2017guv}. Given the symmetries of the spacetime, looking for a scalar field $\phi=\phi\left(r\right)$, the only nonvanishing component of the current is $J^r$. The idea of~\cite{Babichev:2017guv} was to inspect carefully the form of $J^r$, in order to identify which Horndeski functionals enabled a non-trivial scalar field from the equation $J^r=0$ {\it{for spherical symmetry}}. This results in the following pure Horndeski theory with two coupling constants $\eta$ (dimensionless) and $\beta$ (dimension $\text{length}$),
\begin{equation}
G_2= \eta X,\quad G_4 = 1+\beta\sqrt{-X},\quad F_4=0,
\end{equation}
where the $G_2$ term is a canonical kinetic term, and the constant $1$ in $G_4$ corresponds to a pure Einstein-Hilbert term. The BH is homogeneous, $h=f$ in the metric~(\ref{eq:ansatz}), with potential
\begin{equation}
f\left(r\right)=1-\frac{2M}{r}-\frac{\beta^2}{2\eta r^2},
\end{equation}
while the scalar field reads
\begin{equation}
\phi'\left(r\right) = \pm \frac{\sqrt{2}\beta}{\eta r^2\sqrt{f\left(r\right)}}.\label{eq:phi_bcl}
\end{equation}
The previous is solution only if $\eta$ and $\beta$ have the same sign. In the metric, $M$ is an integration constant and is the Arnowitt-Deser-Misner (ADM) mass of the spacetime~\cite{Arnowitt:1961zz}, and also its thermodynamic mass~\cite{Gibbons:1976ue}. The spacetime is clearly asymptotically Schwarzschild at leading order, but differs from the GR result for finite $r$. It presents a curvature singularity at $r=0$. If $\eta>0$, there is a unique horizon, while for $\eta<0$, there are no horizons if $M<\left\lvert\beta\right\rvert/\sqrt{-2\eta}$ (naked singularity) and two horizons if $M>\left\lvert\beta\right\rvert/\sqrt{-2\eta}$. As regards the scalar field, it is real and finite for any $r\geq r_h$, while it becomes imaginary when $f\left(r\right)<0$. Its kinetic term is finite apart from the central singularity,
\begin{equation}
X = -\frac{\beta^2}{\eta^2 r^4},
\end{equation}
and is negative, consistently with the $\sqrt{-X}$ appearing in $G_4$. Note that when we set $M=0$ we do not have flat spacetime but rather a non trivial spacetime. The linear perturbations around this BH have been recently discussed in~\cite{Langlois:2021aji}. 

\subsubsection{Beyond Horndeski black hole}
More recently, the authors of~\cite{Bakopoulos:2022csr} undertook a systematic study of the shift-symmetric beyond Horndeski field equations in spherical symmetry, rewriting them in a way facilitating their integration, in particular for the parity-symmetric case~(\ref{eq:shift-parity}), which permits to identify theories admitting solutions. For instance, the following theory was obtained, displaying again a canonical kinetic term, but also a nonzero beyond Horndeski $F_4$ term,
\begin{equation}
G_2 = \frac{8\eta\beta^2}{\lambda^2}X,\quad G_4 = 1 +4\eta\beta\left(\sqrt{-X}+\beta X\right),\quad F_4 = -\frac{\eta \beta^2}{X},\label{eq:bh_bh}
\end{equation}
with three couplings $\eta$ (dimensionless), $\beta$, $\lambda$, with both $\beta$ and $\lambda$ positive and of dimension $\text{length}$. The metric solution is homogeneous, $h=f$ in~(\ref{eq:ansatz}), where
\begin{equation}
f\left(r\right) = 1-\frac{2M}{r}+\eta\frac{\arctan\left(r/\lambda\right)-\pi/2}{r/\lambda}.\label{eq:arct}
\end{equation}
Again, $M$ is an integration constant and the mass of the solution, which behaves asymptotically as
\begin{equation}
f\left(r\right) = 1-\frac{2M}{r}-\frac{\eta\lambda^2}{r^2}+\mathcal{O}\left(\frac{1}{r^4}\right).
\end{equation}
The scalar field supporting the solution is $\phi=\phi\left(r\right)$ where
\begin{equation}
\phi'\left(r\right) = \pm\frac{1}{\beta\sqrt{2f\left(r\right)}\left(1+\left(r/\lambda\right)^2\right)},\label{eq:phi_bh_arct}
\end{equation}
and is seen to be real and finite for any $r$ greater than the horizon, $r\geq r_h$, and imaginary when $f\left(r\right)<0$, just as for the previous case. The kinetic term is well-defined everywhere, including now at $r=0$,
\begin{equation}
X = -\frac{\lambda^4}{4\beta^2\left(r^2+\lambda^2\right)^2}.\label{eq:x_bh_arct}
\end{equation}
Concerning the horizons, their presence or absence can be understood by looking at the behaviour of the metric near $r=0$,
\begin{equation}
f\left(r\right)=1+\eta-\frac{2M+\pi\eta\lambda/2}{r}-\frac{\eta r^2}{3\lambda^2}+\mathcal{O}\left(r^4\right),
\end{equation}
where appears a threshold mass $M_0 = -\pi\eta\lambda/4$. For $M>M_0$, there is a unique horizon (this is in particular the case for $\eta>0$); for $M<M_0$,  there are two horizons if $M$ is not too small, and zero horizon (naked singularity) for small masses; while for the limit case $M=M_0$, there is zero or one horizon according to $\eta\gtrless -1$, and $f\left(r\right)$ does not diverge at $r=0$ but the spacetime remains singular there because $f\left(0\right)\neq 1$. 

\subsubsection{Non-homogeneous black hole}
Furthermore, the formalism developed in~\cite{Bakopoulos:2022csr} makes it possible to construct non-homogeneous BHs, with $f\neq h$ in~(\ref{eq:ansatz}). This article presents the following example theory admitting such a solution,
\begin{align}
G_2 ={}&{} \frac{8\eta\beta^2}{\lambda^2}\frac{X}{1-\xi^2 X},\quad G_4 = \frac{1 +4\eta\beta\left(\sqrt{-X}+\beta X\right)}{1-\xi^2 X},\label{eq:th_inh_1}\\ 
F_4 ={}&{} \frac{\xi^6X^2-\left(3\xi^2+4\eta\beta^2\right)\xi^2 X-8\eta\beta\xi^2\sqrt{-X}-4\eta\beta^2}{4X\left(1-\xi^2 X\right)^2}.\label{eq:th_inh_2}
\end{align}
As compared to the previously considered theory~(\ref{eq:bh_bh}), there is an additional coupling $\xi$ of dimension $\text{length}$. Although the expression of $F_4$ is largely modified, those of $G_2$ and $G_4$ are only changed by a factor $1/\left(1-\xi^2 X\right)$, which is in fact directly linked to the 'inhomogeneity' of the black hole: the metric function $h\left(r\right)$ which multiplies $-\mathrm{d}t^2$ is as before,
\begin{equation}
h\left(r\right) = 1-\frac{2M}{r}+\eta\frac{\arctan\left(r/\lambda\right)-\pi/2}{r/\lambda},
\end{equation}
see Eq.~(\ref{eq:arct}), while the metric function $f\left(r\right)$ which divides $\mathrm{d}r^2$ becomes
\begin{equation}
f\left(r\right)=\frac{h\left(r\right)}{\left(1-\xi^2 X\right)^2}.\label{eq:inh}
\end{equation}
The scalar field reads
\begin{equation}
\phi'\left(r\right) = \pm\frac{1}{\beta\sqrt{2f\left(r\right)}\left(1+\left(r/\lambda\right)^2\right)},
\end{equation}
which is the same expression in terms of $f\left(r\right)$ as above, Eq.~(\ref{eq:phi_bh_arct}), but $f\left(r\right)$ is now different. Hence, the kinetic term $X=-f\left(r\right)\phi'\left(r\right)^2/2$ is exactly identical to~(\ref{eq:x_bh_arct}), with the behaviour $X=\mathcal{O}\left(1/r^4\right)$ at infinity, which renders the theory~(\ref{eq:th_inh_1}-\ref{eq:th_inh_2}) indistinguishable from the previous one~(\ref{eq:bh_bh}) in this limit, with a canonical kinetic term and Einstein-Hilbert term. The inhomogeneity~(\ref{eq:inh}) is thus asymptotically suppressed with a factor $1/r^4$, but grows significant as $r$ becomes comparable to the coupling $\lambda$. Importantly, the negative sign of $X$ ensures that the denominator in~(\ref{eq:inh}) is always greater than $1$, so in particular never vanishes. In addition, $X$ is bounded everywhere, see~(\ref{eq:x_bh_arct}), therefore, $f\left(r\right)$ vanishes only when $h\left(r\right)$ does, and because $h\left(r\right)$ has the same expression as the previous metric potential~(\ref{eq:arct}), the inhomogeneous spacetime has the same horizons as the previous homogeneous spacetime, with again a singularity at $r=0$.
\subsection{Non-parity symmetric theories}\label{subsec:nonparity}

\subsubsection{Construction of the four-dimensional Einstein-Gauss-Bonnet gravity}
Exact solutions in theories without parity symmetry remained unknown for a long time. This was quite frustrating as interesting numerical solutions had been known for a long time~\cite{Kanti:1995vq} with interesting beyond GR phenomenology. Also for certain theories there were found interesting physical effects such as BH scalarisation~\cite{Doneva:2017bvd}, \cite{Cunha:2019dwb} (for an updated review see \cite{Antoniou:2023gwd}). This changed in 2020, in relation with the quest for a four-dimensional Einstein-Gauss-Bonnet theory of gravity (see~\cite{Fernandes:2022zrq} for a review). The present section presents a concise overview of this search, while the reader mostly interested in the resulting theory and its BH solutions can skip directly to Eq.~(\ref{eq:gen_conf}). 

In order to understand the relevance of the Gauss-Bonnet curvature invariant, one must have in mind the Lovelock theorem~\cite{Lovelock:1971yv} which in essence states: in $D$-dimensional spacetime, the most general action in terms of the metric only and yielding second-order conserved field equations is
\begin{equation}
S = \int\mathrm{d}^Dx\sum_{k=0}^{\lfloor\frac{D-1}{2}\rfloor}\alpha_k\mathcal{R}^{\left(k\right)},\label{eq:lov}
\end{equation} 
where the upper limit of the sum is the floor of $\left(D-1\right)/2$. The $\alpha_k$s are coupling constants, while $\mathcal{R}^{\left(k\right)}$ is a curvature invariant of order $k$, given by
\begin{equation}
\mathcal{R}^{\left(k\right)} = \left(2k\right)!\,\delta^{\mu_1}_{\left[\alpha_1\right.}\delta^{\nu_1}_{\beta_1}\cdots\delta^{\mu_k}_{\alpha_k}\delta^{\nu_k}_{\left.\beta_k\right]}\prod_{i=1}^{k} R^{\alpha_i\beta_i}_{\quad\,\mu_i\nu_i},
\end{equation}
where the brackets denote the usual antisymmetrization. One can normalize the couplings so that $\alpha_1=1$, and call $\alpha_0=-2\Lambda$, then the sum up to $k=2$ reads
\begin{equation}
\sum_{k=0}^{2}\alpha_k\mathcal{R}^{\left(k\right)} = -2\Lambda + R + \alpha_2\mathcal{G},\label{eq:egb}
\end{equation}
where $\mathcal{G}\equiv\mathcal{R}^{\left(2\right)}=R^2-4R_{\mu\nu}R^{\mu\nu}+R_{\mu\nu\rho\sigma}R^{\mu\nu\rho\sigma}$ is known as the Gauss-Bonnet invariant. The Gauss-Bonnet term appears as the first higher-order correction to the Einstein-Hilbert action, and~(\ref{eq:egb}) is usually called Einstein-Gauss-Bonnet gravity. Nevertheless, in $D=4$ dimensions, the sum in~(\ref{eq:lov}) stops at $k=1$ and one recovers GR, because for any $k\geq 3$, $\mathcal{R}^{\left(k\right)}$ vanishes identically, while $\mathcal{R}^{\left(2\right)}=\mathcal{G}$ does not vanish in general, but is a boundary term which does not contribute to the field equations by virtue of the Chern-Gauss-Bonnet theorem\footnote{The Chern-Gauss-Bonnet theorem~\cite{chern} applied to a four-dimensional manifold $\mathcal{M}$ relates its Euler characteristic $\chi\left(\mathcal{M}\right)$ to the integral of the Gauss-Bonnet curvature, $\chi\left(\mathcal{M}\right) = \int_\mathcal{M}\mathcal{G}\,\mathrm{d}\mu/\left(8\pi^2\right)$, where $\mathrm{d}\mu$ is the volume form.}. Only from $D=5$ dimensions and higher does the Gauss-Bonnet term give a non-trivial contribution to Lovelock's action. This is precisely why Lovelock's theorem implies, as argued in the introduction to this chapter, that any modification of gravity in four dimensions must include other degrees of freedom than the two degrees of freedom of the metric. Importantly however, in the framework of $D=4$ scalar-tensor theories, the Gauss-Bonnet term gives non-trivial contributions when non-minimally coupled to the scalar field $\phi$, in actions such as
\begin{equation}
\int\mathrm{d}^4x\sqrt{-g}\Bigl\{R+f\left(\phi\right)\mathcal{G}+\cdots\Bigr\}
\end{equation}
where $f\left(\phi\right)$ is a function of the scalar field\footnote{As regards BH physics, such scalar-tensor Gauss-Bonnet models were known since 1995 to allow for hairy BH solutions~\cite{Kanti:1995vq}, giving then rise to numerous other static~\cite{Sotiriou:2013qea} or spinning~\cite{Kleihaus:2011tg} solutions, all requiring numerical analysis or power series expansions~\cite{Julie:2019sab}. These theories also lead to spontaneous scalarization, again for static~\cite{Doneva:2017bvd} or rotating~\cite{Cunha:2019dwb} BHs.}. In addition to Lovelock's mathematical result, the Gauss-Bonnet invariant is physically motivated by many considerations from string theory, where for instance, 1-loop corrected heterotic string effective action presents terms as $\mathrm{e}^\phi\mathcal{G}$ where $\phi$ is the dilaton~\cite{Zwiebach:1985uq}. 

In this context, one can wonder what is the natural non-trivial generalization of Einstein-Gauss-Bonnet gravity in four dimensions. A first attempt was made by~\cite{Charmousis:2012dw} through a diagonal Kaluza-Klein dimensional reduction~\cite{Overduin:1997sri}. This consists in starting from the $D$-dimensional Einstein-Gauss-Bonnet gravity,
\begin{equation}
\int\mathrm{d}^Dx\sqrt{-g_{(D)}}\Bigl\{R_{(D)}+\hat{\alpha}\mathcal{G}_{(D)}\Bigr\},\label{eq:higherd}
\end{equation}
with coupling $\hat{\alpha}$ and the subscript $\left(D\right)$ stands for $D$-dimensional quantities. Then, the $D$-dimensional metric $\mathrm{d}l_{(D)}^2$ splits into a four-dimensional \textit{spacetime} $\mathrm{d}s_{(4)}^2$ and an $n=D-4$-dimensional \textit{internal space} $\mathrm{d}\widetilde{s}_{(n)}^2$,
\begin{equation}
\mathrm{d}l_{(D)}^2=\mathrm{d}s_{(4)}^2+\mathrm{e}^{-2\phi}\mathrm{d}\widetilde{s}_{(n)}^2,\label{eq:diag}
\end{equation}
and the scalar field $\phi$ appears in the conformal factor of the internal space and depends only on the spacetime coordinates. The authors of~\cite{Charmousis:2012dw} then compute that, up to integration by parts,
\begin{equation}
\int\mathrm{d}^Dx\sqrt{-g_{(D)}}R_{(D)} = T\int\mathrm{d}^4x\sqrt{-g}\,\mathrm{e}^{-n\phi}\Bigl\{R+\widetilde{R}\mathrm{e}^{2\phi}+n\left(n-1\right)\left(\partial\phi\right)^2\Bigr\},\label{eq:r}
\end{equation}
and 
\begin{align}
\int\mathrm{d}^Dx{}&{}\sqrt{-g_{(D)}}\mathcal{G}_{(D)} = T\int\mathrm{d}^4x\sqrt{-g}\,\mathrm{e}^{-n\phi}\Bigl\{\mathcal{G}+\widetilde{\mathcal{G}}\mathrm{e}^{4\phi}\nonumber\\ {}&{}+2\widetilde{R}\mathrm{e}^{2\phi}\Bigl[R+\left(n-2\right)\left(n-3\right)\left(\partial\phi\right)^2\Bigr]-4n\left(n-1\right)G^{\mu\nu}\phi_\mu\phi_\nu\nonumber\\ {}&{}+2n\left(n-1\right)\left(n-2\right)\Box\phi\left(\partial\phi\right)^2-n\left(n-1\right)^2\left(n-2\right)\left(\partial\phi\right)^4\Bigr\},\label{eq:g}
\end{align}
where spacetime quantities of $\mathrm{d}s_{(4)}^2$ are as usual $R$, $\mathcal{G}$, while quantities from the internal space $\mathrm{d}\widetilde{s}_{(n)}^2$ are with tildes, $\widetilde{R}$, $\widetilde{\mathcal{G}}$. One can forget about the global proportionality factor $T$, which is the volume of the internal space, $T = \int\mathrm{d}^nx\sqrt{\widetilde{g}}$. The resulting four-dimensional action is therefore
\begin{align}
S ={}&{} \int\mathrm{d}^4x\sqrt{-g}\,\mathrm{e}^{-n\phi}\Bigl\{R+\widetilde{R}\mathrm{e}^{2\phi}+n\left(n-1\right)\left(\partial\phi\right)^2\nonumber\\{}&{}+\hat{\alpha}\Bigl[\mathcal{G}+\widetilde{\mathcal{G}}\mathrm{e}^{4\phi}+2\widetilde{R}\mathrm{e}^{2\phi}\Bigl(R+\left(n-2\right)\left(n-3\right)\left(\partial\phi\right)^2\Bigr)-4n\left(n-1\right)G^{\mu\nu}\phi_\mu\phi_\nu\nonumber\\ {}&{}+2n\left(n-1\right)\left(n-2\right)\Box\phi\left(\partial\phi\right)^2-n\left(n-1\right)^2\left(n-2\right)\left(\partial\phi\right)^4\Bigr]\Bigr\},\label{eq:egbkk}
\end{align}
and in this action, one can now analytically continue $n$ to be a real parameter. The authors of~\cite{Charmousis:2012dw} were interested in getting a four-dimensional solution descending directly from a $D$-dimensional solution of~(\ref{eq:higherd}), and found that a sufficient condition for this was to take the internal space $\mathrm{d}\widetilde{s}_{(n)}^2$ to be a product of $n/2$ two-spheres of same radius $\rho$~\cite{Dotti:2005rc}, yielding 
\begin{equation}
\widetilde{R} = n/\rho^2,\quad \widetilde{\mathcal{G}} = n\left(n-2\right)/\rho^4.\label{eq:prodspheres}
\end{equation}
This procedure led them to scalar-tensor solutions to action~(\ref{eq:egbkk}) for any $n$, nevertheless, the obtained metric, of the spherically-symmetric form~(\ref{eq:ansatz}) with $h=f$, behaves asymptotically as $f\left(r\right)\approx 1/\left(n+1\right)\left(1-2M/r^{n+1}\right)$, which is not a Newtonian behaviour unless $n=0$, but in this case~(\ref{eq:egbkk}) reduces to pure GR. The way out of this deadlock was first described\footnote{To be precise the renewed interest for constructing a four-dimensional Einstein-Gauss-Bonnet gravity originates from the adventurous and interesting attempt of~\cite{Glavan:2019inb}, despite its shortcomings~\cite{Gurses:2020ofy}.} by~\cite{Lu:2020iav}, which proposed the following procedure: take $n\to 0$ and the coupling $\hat{\alpha}\to \infty$, while maintaining the product $n\hat{\alpha}=\text{const.}\equiv \alpha$, with $\alpha$ a new coupling. In~\cite{Lu:2020iav}, the authors apply this procedure to a maximally-symmetric internal space of curvature $\gamma$, $\widetilde{R}_{abcd} = \gamma\left(g_{ac}g_{bd}-g_{ad}g_{bc}\right)$, implying
\begin{equation}
\widetilde{R} = \gamma n\left(n-1\right),\quad \widetilde{\mathcal{G}} = \gamma^2n\left(n-1\right)\left(n-2\right)\left(n-3\right),\label{eq:maxsym}
\end{equation}
but their prescription in fact applies to any internal space such that $\widetilde{R}\propto n$ and $\widetilde{\mathcal{G}}\propto n$, like~(\ref{eq:prodspheres}), because these terms, multiplied by $\hat{\alpha}$ in~(\ref{eq:egbkk}), then possess a regular limit. The only term of~(\ref{eq:egbkk}) which remains ambiguous in this limit is $\mathrm{e}^{-n\phi}\hat{\alpha}\mathcal{G}$, and it is schematically regularized by expanding the exponential as
\begin{equation}
\mathrm{e}^{-n\phi}\hat{\alpha}\mathcal{G} = \underbrace{\cancel{\hat{\alpha}\mathcal{G}}}_\text{BT} - n\hat{\alpha}\phi\mathcal{G} +\mathcal{O}\left(n^2\right) \underset{n\to 0}{\longrightarrow} -\alpha\phi\mathcal{G}, 
\end{equation}
where in the intermediate step, $\hat{\alpha}\mathcal{G}$ drops out since it is a boundary term (BT) in four dimensions. Very generally, the procedure of~\cite{Lu:2020iav} can thus be applied for any internal space for which the following regularized curvature invariants are well-defined,
\begin{equation}
\widetilde{R}_\text{reg} = \lim_{n\to 0}\frac{\widetilde{R}}{n},\quad \widetilde{\mathcal{G}}_\text{reg} = \lim_{n\to 0}\frac{\widetilde{\mathcal{G}}}{n},\label{eq:curv_reg}
\end{equation}
bringing about the regularized Kaluza-Klein action,
\begin{align}
S = \int\mathrm{d}^4x\sqrt{-g}\Bigl\{R+\alpha\Bigl[-\phi{}&{}\mathcal{G}+\widetilde{\mathcal{G}}_\text{reg}\mathrm{e}^{4\phi}+2\widetilde{R}_\text{reg}\mathrm{e}^{2\phi}\Bigl(R+6\left(\partial\phi\right)^2\Bigr)\nonumber\\ {}&{}+4G^{\mu\nu}\phi_\mu\phi_\nu+4\Box\phi\left(\partial\phi\right)^2+2\left(\partial\phi\right)^4\Bigr]\Bigr\}.\label{eq:kkreg}
\end{align}
Shortly after the publication of~\cite{Lu:2020iav}, two other articles, Refs.~\cite{Fernandes:2020nbq} and~\cite{Hennigar:2020lsl}, came in almost simultaneously with another proposal for regularizing Einstein-Gauss-Bonnet gravity in four dimensions. In this case, the Gauss-Bonnet piece in four dimensions is defined by first considering the $D$-dimensional piece $\int\mathrm{d}^Dx\sqrt{-g}\hat{\alpha}\mathcal{G}$, second, substracting the identical action but for a conformally related metric $\widetilde{g}_{\mu\nu}=\mathrm{e}^{2\phi}g_{\mu\nu}$, then redefining the coupling $\hat{\alpha}=\alpha/\left(D-4\right)$ and finally computing the limit when $D\to 4$. The obtained action is nothing but~(\ref{eq:kkreg}) with $\widetilde{R}_\text{reg}=\widetilde{\mathcal{G}}_\text{reg}=0$, which can thus be seen as a Kaluza-Klein along a flat internal space.

Last but not least, another complementary point of view leads to the regularized Kaluza-Klein action~(\ref{eq:kkreg}): in~\cite{Fernandes:2021dsb}, the author is looking for the most general four-dimensional scalar-tensor action with second order field equations and such that the scalar field equation of motion has conformal symmetry, that is to say, is invariant under $g_{\mu\nu}\to\mathrm{e}^{2\sigma}g_{\mu\nu}$, $\phi\to\phi-\sigma$, with $\sigma$ an arbitrary function on spacetime. Requiring such a symmetry is not so strange at it might first seem. Indeed, the BBMB action, briefly presented in the introduction, see~(\ref{eq:bbmbac}), is usually said to enjoy such a conformal invariance\footnote{More precisely, with the notations of~(\ref{eq:bbmbac}), the invariance is under $g_{\mu\nu}\to\Omega^2g_{\mu\nu}$, $\phi\to\Omega^{-1}\phi$, which is equivalent upon redefining the scalar field $\phi\to\mathrm{e}^\phi$.}, but striclty speaking, this holds true only for the scalar-tensor part $\sqrt{-g}\left(\left(\partial\phi\right)^2/2+R\phi^2/12\right)$, while the pure Einstein-Hilbert term breaks this invariance. Hence, in the BBMB case, the conformal invariance holds only at the level of the scalar field equation of motion, and~\cite{Fernandes:2021dsb} questions what is the most general scalar-tensor action with this property\footnote{and, as already said, with second order field equations, otherwise, most general theories are obtained, see~\cite{Ayon-Beato:2023bzp}.}. The result is
\begin{align}
S =\int\mathrm{d}^4x\sqrt{-g}{}&{}\Bigl\{R-2\lambda\mathrm{e}^{4\phi}-\beta\mathrm{e}^{2\phi}\Bigl[R+6\left(\partial\phi\right)^2\Bigr]\nonumber\\ {}&{}+\alpha\Bigl[-\phi\mathcal{G}+4G^{\mu\nu}\phi_\mu\phi_\nu+4\Box\phi\left(\partial\phi\right)^2+2\left(\partial\phi\right)^4\Bigr]\Bigr\},\label{eq:gen_conf}
\end{align}
with three independent couplings $\lambda$, $\beta$ and $\alpha$. This is the same action as~(\ref{eq:kkreg}) with
\begin{equation}
2\lambda = -\alpha\widetilde{\mathcal{G}}_\text{reg},\quad \beta = -2\alpha\widetilde{R}_\text{reg}.\label{eq:link}
\end{equation}
Due to its \textit{generalized conformal invariance}, the field equations of action~(\ref{eq:gen_conf}) combine in the following way,
\begin{equation}
2g^{\mu\nu}\frac{\delta S}{\delta g^{\mu\nu}}+\frac{\delta S}{\delta\phi} = R+\frac{\alpha}{2}\mathcal{G}.
\end{equation}
Hence, any scalar-tensor solution to the field equations of~(\ref{eq:gen_conf}) must satisfy the purely geometric equation $R+\alpha\mathcal{G}/2=0$. This reproduces exactly the trace of the field equations for the purely metric Einstein-Gauss-Bonnet gravity in $D$ dimensions, $\int\mathrm{d}^Dx\sqrt{-g}\Bigl\{R+\alpha\mathcal{G}\Bigr\}$. Given the previously detailed construction and properties of action~(\ref{eq:gen_conf}), it can be called the four-dimensional Einstein-Gauss-Bonnet gravity. As advertised, it is a non-parity symmetric Horndeski theory~(\ref{eq:horndeski_lag}), with functions
\begin{align}
G_2 ={}&{} -2\lambda\mathrm{e}^{4\phi}+12\beta\mathrm{e}^{2\phi}X+8\alpha X^2,\quad G_3 = 8\alpha X,\nonumber\\ G_4 ={}&{} 1-\beta\mathrm{e}^{2\phi}+4\alpha X,\quad G_5 = 4\alpha\log\left\lvert X\right\rvert.\label{eq:horn_conf}
\end{align}
The theory acquires shift-symmetry when $\lambda=\beta=0$, the $\phi\mathcal{G}$ term being shift-symmetric because $\mathcal{G}$ is a boundary term. Note also that the $\mathrm{e}^{2\phi}X$ term in $G_2$ can be transformed into a canonical kinetic term for the scalar field $\Phi\equiv \mathrm{e}^\phi$, but we will continue working with $\phi$. This is important as scalar field perturbations can be affected by the absence of the lowest order term in the action. Let us now present the various exact, asymptotically flat BH solutions to action~(\ref{eq:gen_conf}), of the static, spherically symmetric, homogeneous form,
\begin{equation}
\mathrm{d}s^2 = -f\left(r\right)\mathrm{d}t^2+\frac{\mathrm{d}r^2}{f\left(r\right)}+r^2\mathrm{d}\Omega^2.\label{eq:hom_ans}
\end{equation}
We will classify these solutions according to the {\it{geometric characteristics of the internal space}} as it appears in the Kaluza-Klein picture~(\ref{eq:kkreg}). Importantly, they all correspond to different relative values of the couplings $\lambda$, $\beta$ and $\alpha$, thus they are solutions to different theories.

\subsubsection{Maximally-symmetric internal space}
The first solution was obtained by~\cite{Lu:2020iav}, following their Kaluza-Klein regularization procedure along a maximally-symmetric internal space~(\ref{eq:maxsym}). Using~(\ref{eq:curv_reg}) and~(\ref{eq:link}), this corresponds to $\lambda = 3\beta^2/\left(4\alpha\right)$. The metric is
\begin{equation}
f\left(r\right) = 1+\frac{r^2}{2\alpha}\left(1\pm\sqrt{1+\frac{8\alpha M}{r^3}}\right).\label{eq:met_lu_pang}
\end{equation}
Very interestingly, this profile can be seen as the continuation to $D\to 4$ of the spherically-symmetric Boulware-Deser solution of pure metric Einstein-Gauss-Bonnet gravity in $D\geq 5$~\cite{Boulware:1985wk} (see also~\cite{Glavan:2019inb}). The $\pm$ in front of the square root comes from the quadratic order of the field equations and appears for all solutions to action~(\ref{eq:gen_conf}), but only the $-$ branch is asymptotically flat and reduces to GR in the limit of vanishing coupling $\alpha$, therefore, the $+$ branch is always omitted from now on. The integration constant $M$ is the ADM and thermodynamic mass of the spacetime, which behaves asymptotically as
\begin{equation}
f\left(r\right) = 1-\frac{2M}{r}+\frac{4\alpha M^2}{r^4}+\mathcal{O}\left(\frac{1}{r^7}\right).\label{eq:asymp_lu_pang}
\end{equation}
The scalar field is radial and reads\footnote{The divergence of $\phi\left(r\right)$ as $r\to\infty$ is irrelevant as it can be eliminated by field redefinition. Indeed the redefined scalar $\Phi=\mathrm{e}^\phi$ already mentioned between~(\ref{eq:horn_conf}) and~(\ref{eq:hom_ans}), vanishes as $r\to\infty$.\label{f:redef_scal}}
\begin{equation}
\phi\left(r\right) = \log\left(\frac{\sqrt{-2\alpha/\beta}}{r}\right)-\log\left(\sigma\left(c\pm\int\frac{\mathrm{d}r}{r\sqrt{\left\lvert f\left(r\right)\right\rvert}}\right)\right),\label{eq:phi_lu_pang}
\end{equation}
where $c$ is an arbitrary integration constant and $\sigma$ is a function whose expression depends on the sign of $f\left(r\right)$: if $f\left(r\right)\geq 0$, $\sigma = \cosh$, while if $f\left(r\right)<0$, $\sigma = \left\lvert\cos\right\rvert$. Therefore the scalar field when $\alpha \beta<0$ is well-defined when $f\left(r\right)\geq 0$ (in particular outside the event horizon), and even when $f\left(r\right)<0$ apart on a set of null measure. For $\alpha \beta>0$ the solution can be adequately redefined outside the event horizon but is imaginary once $f<0$. The horizons of~(\ref{eq:met_lu_pang}) depend on the sign of $\alpha$. If $\alpha>0$, there are two possible horizons at
\begin{equation}
r_{\pm} = M\pm\sqrt{M^2-\alpha},\label{eq:rpm}
\end{equation}
in other words, the spacetime has no horizons if $M< \sqrt{\alpha}$ and has two horizons otherwise. For $\alpha<0$, the square root in~(\ref{eq:met_lu_pang}) becomes ill-defined before reaching $r_+$ if the mass is too small (naked singularity), while if the mass is sufficiently large, there is indeed a horizon at $r_+$ but no inner horizon since the square root becomes ill-defined. Ref.~\cite{Charmousis:2021npl} proved that the expression~(\ref{eq:rpm}) for the horizon would strongly constrain the negative values of $\alpha$ if there were a Birkhoff-like theorem for the considered theory. In this case, an atomic nucleus of radius $R$ would produce the gravitational field~(\ref{eq:met_lu_pang}), and is not a BH so verifies $r_+<R$, yielding $-10^{-30}\,\text{m}^2<\alpha<0$. When $\alpha>0$, the metric is well-defined up to $r=0$, where it behaves as
\begin{equation}
f\left(r\right)=1-\sqrt{\frac{2Mr}{\alpha}}+\frac{r^2}{2\alpha}+\mathcal{O}\left(r^{7/2}\right).\label{eq:lu_pang_zero}
\end{equation}
Although $f\left(r\right)$ is well-defined at $r=0$, the curvature invariants diverge there (they would not if $f\left(r\right)=1+\mathcal{O}\left(r^2\right)$, which is the case only for $M=0$ but this reduces to flat spacetime), so the spacetime is singular at $r=0$.

\subsubsection{Flat internal space}
The second solution was presented by~\cite{Hennigar:2020lsl}, after their regularization which corresponds to a flat internal space and therefore to $\lambda=\beta=0$, see~(\ref{eq:link}). The metric potential $f\left(r\right)$ is exactly the same as~(\ref{eq:met_lu_pang}), while the radial scalar field is now given by
\begin{equation}
\phi\left(r\right)=\int\frac{\pm 1-\sqrt{f\left(r\right)}}{r\sqrt{f\left(r\right)}}\mathrm{d}r\label{eq:phi_mann}
\end{equation}
up to a global additive constant since this particular theory is shift-symmetric. The scalar field is well-defined for $r\geq r_+$ but becomes imaginary when $f\left(r\right)<0$. The kinetic term itself, whose regularity is of interest since the theory is shift-symmetric, is imaginary for $f\left(r\right)<0$,
\begin{equation}
X = -\frac{\left(\pm 1-\sqrt{f\left(r\right)}\right)^2}{2r^2}.
\end{equation}
This unpleasant feature was cured in~\cite{Charmousis:2021npl}, by adding a linear time dependence to the scalar field. It now reads
\begin{equation}
\phi = qt + \int\frac{\pm\sqrt{q^2r^2+f\left(r\right)}-f\left(r\right)}{rf\left(r\right)}\mathrm{d}r,\label{eq:cured}
\end{equation}
while the metric remains unchanged and does not feel the influence of the scalar hair $q$, which is an arbitrary integration constant. As explained in the section regarding stealth solutions, the regularity of the scalar field at the horizon $r_+$ is manifest in horizon-crossing coordinates, since $\phi\left(v,r=r_+\right)=qv+\text{const.}$ or $\phi\left(u,r=r_+\right)=qu+\text{const.}$ depending on the sign of $q$ and the choice of $\pm$ in~(\ref{eq:cured}), where $v=t+r^\star$, $u=t-r^\star$ and $r^\star = \int\mathrm{d}r/f\left(r\right)$. Moreover, since $f\left(r\right)$ as given by~(\ref{eq:met_lu_pang}) is bounded from below (with a minimum depending on the mass $M$), there always exists a range of the integration constant $q$ such that the scalar field is well-defined in the whole spacetime, and so is the kinetic term,
\begin{equation}
X = \frac{\pm 2\sqrt{q^2r^2+f\left(r\right)}-1-f\left(r\right)}{2r^2},
\end{equation}
apart from the central singularity at $r=0$ (this is still true even if $M=0$). Note that $X$ here is not constant unlike the stealth solutions we studied before. As a result disformal transformations of this solution can lead, in certain cases, to different spacetimes, such as traversable regular wormhole solutions~\cite{Bakopoulos:2021liw}. This theory also involves interesting neutron star solutions with quite distinct properties from GR such as a universal absence of mass gap in between neutron stars and black holes~\cite{Charmousis:2021npl}.

\subsubsection{Internal space as a product of two spheres}
The third solution was described by~\cite{Fernandes:2021dsb}, and in fact corresponds to the Kaluza-Klein regularization for an internal space which is a product of $n/2$ two-spheres of identical radius, see~(\ref{eq:prodspheres}). Eqs.~(\ref{eq:curv_reg}) and~(\ref{eq:link}) imply that $\lambda = \beta^2/\left(4\alpha\right)$, and the metric is
\begin{equation}
f\left(r\right) = 1+\frac{r^2}{2\alpha}\left(1-\sqrt{1+\frac{8\alpha M}{r^3}+\frac{8\alpha^2}{r^4}}\right),\label{eq:prod_sphere_sol}
\end{equation}
$M$ being again the mass. The asymptotic departure from Schwarzschild gets larger than in~(\ref{eq:asymp_lu_pang}), since now,
\begin{equation}
f\left(r\right)=1-\frac{2M}{r}-\frac{2\alpha}{r^2}+\mathcal{O}\left(\frac{1}{r^4}\right).
\end{equation}
In fact, this solution directly descends from a higher-dimensional solution~\cite{Dotti:2005rc} through the diagonal Kaluza-Klein reduction~(\ref{eq:diag})~\cite{Charmousis:2012dw}, and the scalar field, whose exponential $\mathrm{e}^{-2\phi}$ multiplies the two-spheres of the internal space, has consistently a very simple expression\footnote{Footnote~\ref{f:redef_scal} explains how to get rid of the divergence of $\phi\left(r\right)$ as $r\to\infty$.},
\begin{equation}
\phi\left(r\right)=\log\left(\frac{\sqrt{-2\alpha/\beta}}{r}\right).\label{eq:log_scal}
\end{equation}
The candidate event horizon is now located at
\begin{equation}
r_+ = M+\sqrt{M^2+\alpha}.
\end{equation}
For $\alpha>0$, there is always a unique horizon at $r_+$, while for $\alpha<0$, the square root might become ill-defined: it turns out that there is no horizon for small masses, and a unique horizon at $r_+$ for sufficiently large masses. This was studied in detail in~\cite{Babichev:2022awg}. Also, the same kind of Birkhoff conjecture as explained between~(\ref{eq:rpm}) and~(\ref{eq:lu_pang_zero}) now severly constrains the positive values of $\alpha$ for the present theory, $0<\alpha<10^{-30}\,\text{m}^2$. When the metric reaches $r=0$ with no problems in the square root, it behaves there as
\begin{equation}
f\left(r\right) = 1-\text{sgn}\left(\alpha\right)\sqrt{2}-\frac{Mr}{\sqrt{2}\left\lvert\alpha\right\rvert}+\mathcal{O}\left(r^2\right).
\end{equation}
Again, $f\left(r\right)$ is well-defined at $r=0$, but not sufficiently regular, so there is a curvature singularity at $r=0$. A major difference with the previous spacetime~(\ref{eq:met_lu_pang}) is that~(\ref{eq:prod_sphere_sol}) does not reduce to flat spacetime when $M=0$, but is rather a naked singularity at $r=0$. This is due to the $8\alpha^2/r^4$ term in~(\ref{eq:prod_sphere_sol}), which ultimately comes from the higher-dimensional origin of the solution with a horizon of non-trivial topology (product of two-spheres)~\cite{Charmousis:2012dw},~\cite{Dotti:2005rc}. Note that disformal transformations of such solutions can lead to wormhole solutions~\cite{Babichev:2022awg}.

As a summary, the four-dimensional Einstein-Gauss-Bonnet gravity~(\ref{eq:gen_conf}) admits three distinct scalar-tensor BH solutions, for three distinct theories with respectively $\lambda=3\beta^2/\left(4\alpha\right)$, $\lambda=\beta=0$ and $\lambda=\beta^2/\left(4\alpha\right)$. Ref.~\cite{Fernandes:2021dsb} in fact showed that these are the only ones\footnote{More precisely, Ref.~\cite{Fernandes:2021dsb} also finds a solution with constant scalar field, but we do not mention it because the field equations become partly degenerate at this constant value (strong coupling). This constant scalar even formally allows for a non-perturbative rotating solution~\cite{Fernandes:2023vux}, but degeneracy is manifest since this solution is parameterized by two arbitrary functions of $\theta$ which are not fixed by the field equations.} with the homogeneous\footnote{Furthermore, for the shift-symmetric case $\lambda=\beta=0$, unicity of the static BH was proven~\cite{Fernandes:2021ysi}, in other words, non-homogeneous BHs do not exist in this case.} ansatz~(\ref{eq:hom_ans}). Although each of the obtained theory can be mapped to the regularized Kaluza-Klein action~(\ref{eq:kkreg}) with different internal space (respectively maximally-symmetric, flat, and product of two-spheres), as regards the solutions, only the product of two-spheres case directly descends from a solution to Einstein-Gauss-Bonnet gravity in $D$ dimensions through the diagonal Kaluza-Klein reduction~(\ref{eq:diag}), while a hypothetical Kaluza-Klein origin for the other two solutions remains unclear. Indeed, their metric function~(\ref{eq:met_lu_pang}) generalizes the higher-dimensional Boulware-Deser solution~\cite{Boulware:1985wk}, but their scalar field~(\ref{eq:phi_lu_pang}) or~(\ref{eq:phi_mann}) does not fit into the diagonal Kaluza-Klein reduction picture~(\ref{eq:diag}).

In a nutshell, the search for a four-dimensional Einstein-Gauss-Bonnet gravity provided as a byproduct quite involved Horndeski theories, without parity nor shift symmetry in general and which, quite remarkably, admit exact BH solutions. Nevertheless, the formulation~(\ref{eq:gen_conf}) of~\cite{Fernandes:2021dsb} shows that the resulting theory still admits a kind of symmetry, the generalized conformal invariance. The next paragraph provides an example of a Horndeski theory generalizing~(\ref{eq:gen_conf}) with terms breaking this generalized conformal invariance, and still admitting exact solutions.

\subsubsection{A Horndeski theory without any symmetry}
The authors of~\cite{Babichev:2023dhs} constructed this theory by first considering an action similar to~(\ref{eq:gen_conf}), but with generalized potentials in front of each term (for instance $\phi\mathcal{G}\to V\left(\phi\right)\mathcal{G}$, etc.), and then imposing that the resulting theory allow for a logarithmic scalar field similar to~(\ref{eq:log_scal}). The obtained action reads
\begin{align}
S = {}&{}\int\mathrm{d}^4x\sqrt{-g}\Bigl\{
R-2\lambda_4\mathrm{e}^{4\phi}-2\lambda_5\mathrm{e}^{5\phi}-
\beta_4\mathrm{e}^{2\phi}\left(R+6\left(\partial\phi\right)^2\right)\nonumber\\{}&{}-
\beta_5\mathrm{e}^{3\phi}\left(R+12\left(\partial\phi\right)^2\right)
-\alpha_4\left(\phi\mathcal{G}-4G^{\mu\nu}\phi_\mu\phi_\nu-
4\Box\phi\left(\partial\phi\right)^2-2\left(\partial\phi\right)^4\right)\nonumber\\{}&{}-\alpha_5
\mathrm{e}^{\phi}\left(\mathcal{G}-8G^{\mu\nu}\phi_\mu\phi_\nu-12\Box\phi
\left(\partial\phi\right)^2-12\left(\partial\phi\right)^4\right)\Bigr\},
\label{eq:complete}
\end{align}
with six coupling constants $\lambda_4$, $\lambda_5$, $\beta_4$, $\beta_5$, $\alpha_4$ and $\alpha_5$. The part with $\lambda_4$, $\beta_4$ and $\alpha_4$ is identical to the four-dimensional Einstein-Gauss-Bonnet gravity~(\ref{eq:gen_conf}) with the generalized conformal invariance. On the other hand, the couplings with $\lambda_5$, $\beta_5$ and $\alpha_5$ would correspond to a conformally-invariant scalar field in \textit{five} dimensions~\cite{Oliva:2011np}: the five-dimensional action
\begin{align}
\int\mathrm{d}^5x\sqrt{-g}\Bigl\{2\lambda_5{}&{}\mathrm{e}^{5\phi}+\beta_5\mathrm{e}^{3\phi}\left(R+12\left(\partial\phi\right)^2\right)\nonumber\\{}&{}+\alpha_5
\mathrm{e}^{\phi}\left(\mathcal{G}-8G^{\mu\nu}\phi_\mu\phi_\nu-12\Box\phi
\left(\partial\phi\right)^2-12\left(\partial\phi\right)^4\right)\Bigr\}
\end{align}
is invariant under $g_{\mu\nu}\to\mathrm{e}^{2\sigma}g_{\mu\nu}$, $\phi\to\phi-\sigma$. Of course, this invariance does not hold in the four-dimensional action~(\ref{eq:complete}). It is hard to understand why a Lagrangian with interesting properties in five dimensions would fit in a four-dimensional action and give exact solutions (which we discuss in a moment). A possible interpretation, which was not given in~\cite{Babichev:2023dhs}, emerges by looking at the Kaluza-Klein reduction of the Gauss-Bonnet action into four dimensions, Eq.~(\ref{eq:g}). In this expression, $n$ initially corresponds to the dimension of the internal space, but is analytically continued to the real axis, and upon taking $n=-1$, the obtained terms are exactly the $\lambda_5$, $\beta_5$ and $\alpha_5$ terms of~(\ref{eq:complete}), provided the curvature invariants of the internal space are related to the couplings as
\begin{equation}
2\lambda_5=-\alpha_5\widetilde{\mathcal{G}},\quad \beta_5=-2\alpha_5\widetilde{R}.
\end{equation}
This interpretation at least shows how this five-dimensional conformal scalar field can appear in four dimensions, but the negative dimension $n=-1$ of the internal space is mysterious, and it remains unclear why this dimension is special and gives exact BH solutions (which is not the case for other negative values of $n$). Action~(\ref{eq:complete}) belongs to the Horndeski class with complicated functions $G_k$, $2\leq k\leq 5$, which can be found in~\cite{Babichev:2023dhs}.
There exists two solutions of the static, spherically-symmetric form~(\ref{eq:hom_ans}), corresponding to different values of the couplings, and therefore to different theories. The first solution requires
\begin{equation}
\lambda_4 = \frac{\beta_4^2}{4\alpha_4},\quad \lambda_5 =
\frac{9\beta_5^2}{20\alpha_5},\quad\frac{\beta_5}{\beta_4} =
\frac{2\alpha_5}{3\alpha_4}
\end{equation}
and reads\footnote{with again a $\pm$ in front of the square root, where the $+$ branch is forgotten because not asymptotically flat.}
\begin{align}
f\left(r\right) = {} & 1 + \frac{2\alpha_5\eta}{3\alpha_4r}
+\frac{r^2}{2\alpha_4}\left[1-
\sqrt{\left(1+\frac{4\alpha_5\eta}{3r^3}\right)^2+4\alpha_4\left(\frac{2
M}{r^3}+\frac{2\alpha_4}{r^4}+\frac{8\alpha_5\eta}{5r^5}\right)}\right],\nonumber\\ 
\phi={}&\ln\left(\frac{\eta}{r}\right),\quad \eta \equiv \sqrt{\frac{-2\alpha_4}{\beta_4}}.\label{eq:sol_1}
\end{align}
This spacetime, with mass $M$, is a generalization of the above one~(\ref{eq:prod_sphere_sol}), with deviation parameterized by the product $\alpha_5\eta$ and of order $\mathcal{O}\left(1/r^3\right)$ at infinity. As for this previous solution, the $M=0$ case is not a flat spacetime. $r=0$ is a curvature singularity, with either $f\left(0\right)=-1/5$ for $\alpha_5>0$ or $\lim_{r\to 0}f\left(r\right)=\infty$ if $\alpha_5<0$, but of course, there might be a curvature singularity before $r=0$ if the square root becomes ill-defined.

The second solution exists for
\begin{equation}
\lambda_4=\beta_4=\alpha_4=0,\quad
\lambda_5=\frac{9\beta_5^2}{20\alpha_5},
\end{equation}
which corresponds to the pure $n=-1$ Kaluza-Klein described above, with a metric function and a scalar field given by
\begin{equation}
f\left(r\right)=\frac{1}{1+\frac{4\alpha_5\eta}{3r^3}}\left[1-\frac{2M}{r}-\frac{4\alpha_5\eta}{15
r^3}\right],\quad \phi=\ln\left(\frac{\eta}{r}\right),\quad \eta \equiv
2\sqrt{\frac{-\alpha_5}{3\beta_5}}.
\label{eq:sol_2}
\end{equation}
It is a simple deviation from Schwarzschild, parameterized by the product $\alpha_5\eta$, and of order $\mathcal{O}\left(1/r^3\right)$ at infinity. If $\alpha_5>0$, the curvature singularity is at $r=0$ with $f\left(0\right)=-1/5$, while if $\alpha_5<0$, the singularity is at $r = \left(-4\alpha_5\eta/3\right)^{1/3}$. Note that for both solutions~(\ref{eq:sol_1}) and~(\ref{eq:sol_2}), the candidate event horizon $r_h$ is root of a cubic equation,
\begin{equation}
15r_h^3-30Mr_h^2-15\alpha_4r_h-4\alpha_5\eta=0,
\end{equation}
with $\alpha_4=0$ for the second case, and the precise existence or not of the horizon depends on the various signs of the couplings, see details in~\cite{Babichev:2023dhs}.

\section{Summary}
Higher-order scalar-tensor theories admit a number of exact asymptotically flat BH solutions in four dimensions, of which this chapter has reviewed the most relevant ones. A first step was accomplished by showing the possibility for stealth solutions (section~\ref{sec:stealth}), where the metric is Schwarzschild or Kerr, but the scalar field is non-trivial and thus modifies the perturbation theory around these GR backgrounds. The stealth Kerr solution even stands as an interesting starting point for generating, by a disformal transformation, the disformed Kerr metric~(\ref{eq:disf_kerr}), which displays new properties as compared to its GR counterpart, in particular due to its non-circularity. However, disformal transformation from a seed stealth solution is by no means the unique way of obtaining non-stealth solutions, that is to say, solutions for which the spacetime differs from the Schwarzschild or Kerr metrics predicted by GR: section~\ref{sec:nonstealth} presents a number of such non-stealth solutions, all of them with the static, spherically-symmetric form~(\ref{eq:ansatz}). The less difficult case of parity and shift-symmetric theories is dealt with in subsection~\ref{subsec:parity}, with three different solutions, two of them being homogeneous (i.e. $g_{tt}=-g^{rr}$) and for which the associated (beyond) Horndeski theories admit a canonical kinetic term, while the third one is non-homogeneous. Then, subsection~\ref{subsec:nonparity} studies non-parity symmetric theories: a great progress was recently achieved in the course of the search for a four-dimensional Einstein-Gauss-Bonnet gravity, which appears as a Horndeski theory enjoying a generalized conformal invariance, see~(\ref{eq:gen_conf}), in relation with the regularized Kaluza-Klein action of higher-dimensional Einstein-Gauss-Bonnet gravity, dimensionally reduced along an internal space of dimension $n\to 0$, see~(\ref{eq:kkreg}). This leads to three scalar-tensor BH solutions, depending on the internal space geometry (maximally-symmetric, flat, or product of two-spheres). Finally, a generalization~(\ref{eq:complete}) of~(\ref{eq:gen_conf}), with additional terms breaking the generalized conformal invariance and possibly linked to a Kaluza-Klein procedure, completes the picture with two more solutions.

Apart from the solutions we have depicted here, DHOST theories have been shown to host exotic solutions such as regular black holes. These solutions are constructed without the need of fine-tuning, where the solution characteristics end up appearing in the coupling constants of the theory in question. These solutions have an inner and outer event horizon with a regular de Sitter central core~\cite{Babichev:2020qpr} and host a large number of regular black hole geometries such as the Hayward solution~\cite{Hayward:2005gi} for example. Quite interestingly, these solutions appear to exist away from beyond Horndeski theories; i.e., only in pure DHOST theories. This may or not be due to the construction method, an extension of the well-known Kerr-Schild solution generating trick. Neutron star solutions have been found in the context of the four dimensional Einstein Gauss Bonnet theory which have interesting and distinct properties from GR~\cite{Charmousis:2021npl}. For example there is no mass gap in between neutron star solutions and black holes and as such neutron stars can be more massive in accord to recent GW observables such as GW190814~\cite{LIGOScientific:2020zkf}. Recently, smooth traversable wormholes have been found as purely vacuum modified gravity solutions in beyond Horndeski theories~\cite{Bakopoulos:2021liw,Babichev:2022awg}. These solutions are obtained starting from Horndeski theory and going via disformal transformation to a beyond Horndeski theory. Although there has been considerable recent progress in the field, the greatest challenge remains: finding explicit stationary solutions. We discussed the disformal Kerr metric and its black hole properties~\cite{Anson:2020trg} which originates from a simple stealth Kerr metric. Very recently an intriguing stationary metric was also proposed~\cite{Fernandes:2023vux}, which originates from semi-classical gravity and is more akin to the theories we studied in section~\ref{sec:nonstealth} (rather than those of section~\ref{sec:stealth} involving stealth metrics). Although the origin of both solutions is distinctively different, both metrics exhibit non circularity, very seldom considered in numerical and ad hoc beyond Kerr constructions (see for example \cite{Johannsen:2013rqa} and references within). It is paramount therefore to find such solutions and their properties and figure out what they signify on the observational front. These are some of the important next steps in the field of explicit compact object solutions beyond GR.

\end{document}